\documentclass[%
reprint,
amsmath,amssymb,
aip, 
apl
]{revtex4-2}

\usepackage{graphicx}
\usepackage{dcolumn}
\usepackage{bm}
\usepackage{wasysym}
\usepackage{comment}
\usepackage{pgfplots}
\usepackage{color,soul}

\usepackage{lipsum} 
\usepackage{tikz}
\usetikzlibrary{positioning}
\newcommand{\tikzcircle}[2][black,fill=white]{\tikz[baseline=-0.5ex]\draw[#1,radius=#2] (0,0) circle ;}%

\usepackage{chemformula}

\usepackage{notes2bib}

\usepackage{bibentry}

\makeatletter
\let\@fnsymbol\@fnsymbol@latex
\@booleanfalse\altaffilletter@sw
\makeatother

\begin{document}

\preprint{XXX/XXXX}

\title{ATP-induced reconfiguration of the micro-viscoelasticity of cardiac and skeletal myosin solutions}

\author{Pablo Dom\'inguez-Garc\'ia}
 \affiliation{Dep. F\'{i}sica Interdisciplinar, Universidad Nacional de Educaci\'{o}n a Distancia (UNED), Madrid 28040, Spain.} 

\author{Jose R. Pinto} 
\affiliation{Department of Biomedical Sciences, Florida State University College of Medicine, Florida FL 32304, USA.}

\author{Ana Akrap}
\affiliation{Department of Physics, University of Fribourg, Fribourg CH-1700, Switzerland. }
\email[Corresponding author: ]{ana.akrap@unifr.ch}

\author{Sylvia Jeney}
\affiliation{Department of Physics, University of Fribourg, Fribourg CH-1700, Switzerland. }

\begin{abstract} 
We study the high-frequency micro-mechanical response of suspensions composed by cardiac and skeletal muscle myosin by optical trapping interferometry.
We observe that in low ionic strength solutions
upon the addition of magnesium adenosine triphosphate (\ch{MgATP^2-}), 
myosin suspensions radically change their micro-mechanics properties, 
generating a viscoelastic fluid characterized by a complex modulus similar to a suspension of worm-like micelles. 
This transduction of energy, from chemical to mechanical, may be related to the relaxed states of myosin, which regulate muscle contractility and can be involved in the etiology of many myopathies.
Within an analogous generic mechanical response, cardiac and skeletal myosin suspensions provide different stress relaxation times, elastic modulus values, and characteristic lengths. 
These discrepancies probably rely on the dissimilar physiological functions of cardiac and skeletal muscle, on the different MgATPase hydrolysis rates of cardiac and skeletal myosins, and on the observed distinct cooperative behavior of their myosin heads in the super-relaxed state.
\textit{In vitro} studies like these allow to understand the foundations of muscle cells mechanics on the micro-scale, and may contribute to the engineering of biological materials whose micro-mechanics can be activated by energy regulators.

\end{abstract}

\maketitle

Myosin is a molecular machine whose isoforms are involved in different processes of eukaryotic cells, such as cell division and movement, or intracellular transport \cite{Kovacs2013}. 
This molecule is well-known because it drives muscle contraction
by binding to F-actin with \ch{Mg^2+}-ATPase activity \cite{Korn1988}. 
Importantly, myosin in its relaxed state, specially from cardiac muscle, can be related to cardioprotective mechanisms \cite{Hooijman2011}.   
A striated muscle myosin II molecule consists of three main structural components: two globular heads (subfragment 1) that can hydrolyze ATP and bind to actin; a double-stranded coiled-coil $\alpha$-helical segment (subfragment 2) that lacks ATPase activity; and a rigid, rod-like coiled-coil section known as light meromyosin, which tends to aggregate and form thick filaments of myosin molecules (see schematics in Refs \cite{Highsmith1981,Chen2024}).

During muscle contraction, the head structure of these molecules is considered the main
part of the protein responsible of the transduction of energy, from chemical to mechanical \cite{Harrington1990}, 
and the time that a myosin head needs to hydrolyze is related to muscle thermogenesis and metabolic activity \cite{Stewart2010}.
Myosin molecules are contained in the thick filament of the striated sacromere, while the thin filament contains actin and regulatory proteins (tropomyosin and troponin).
The interaction of thick and thin filaments generates a phenomenon named super-precipitation \cite{Sekine1966,Hayashi1975}, which has been considered as the \textit{in vitro} contraction \cite{Ebashi1969}. 
When it comes to \textit{in vitro} experiments, studies on actomyosin (AM) solutions traditionally provide simple models for exploring the interactions between actin and myosin, and the mechanical properties of AM networks \cite{Humphrey2002, Koenderink2009, SoareseSilva2011}.

Sarcomeric myosin molecules aggregate leading to filament formation at low ionic strength, e.g., for \ch{KCl} concentrations below $0.6\,$M \cite{Kaminer1966}.
Hence, it is not possible to study monomeric myosin in solution under physiological conditions \cite{Hvidt1984}.
The synthetic myosin filaments have a diameter of $16\,$nm, and a length of about $1.6\,\mu$m, but these quantities vary with solvent conditions \cite{Kaminer1966}, such as the salt content in the solution, and length and width of the filaments decrease when increasing KCl concentration \cite{Pepe1982, Saito1994}. 
The myosin minifilaments \cite{Niederman1975} can be very uniform in size, with a bipolar structure, 
a length of $\sim 0.3-0.4\,\mu$m \cite{Reisler1980, Trybus1987} and a diameter of about $8\,$nm.

The observed rheological properties of dilute myosin solutions are typical of semi-flexible structures and allowed to deduce mechanical properties of the myosin molecule \cite{Rosser1978,Hvidt1982}. 
While the rod domain of the protein has been considered semi-flexible, a second flexible region appears where the heads attach to the tail, which allows the heads to move freely in solution \cite{Curry1991}. 
This section links the heads to the thick filament when adding ATP, creating cross-bridges to bound to actin. 
At this point, some mechanism should temporarily store the energy before the movement of the filaments during muscle contraction is activated, and the elasticity of the cross-bridge is a potential candidate \cite{Highsmith1981}. 

In this work, we study the micro-mechanical properties of solutions of striated muscle myosins, i.e. cardiac and skeletal, to analyze the changes on their micro-viscoelasticity in the presence or absence of \ch{MgATP^2-} without actin. 
Here, we search for the mechanical storing of chemical energy in the myosins network under the effect of ATP, something already observed for actin filaments  \cite{Janmey1990}.
In fact, the state where the myosin heads action is blocked by molecular switches is called the relaxed state \cite{Woodhead2005}. 
An additional relaxed state, named super-relaxed state, with a very low metabolic rate \cite{Hooijman2011}, has been detected for both cardiac and skeletal myosin. 
It has been proposed that the structural base for this state is the autoinhibitory interactions between the heads \cite{Chu2021,Craig2022,Chen2024}.  

\begin{figure}
\begin{center}
\includegraphics[scale=1]{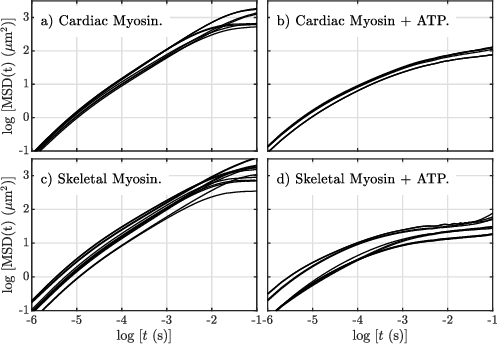}
\end{center}
\caption{\label{fig1}Log-log plots of the one-dimensional mean-square displacements (MSD) for  melamine resin trapped beads using the weakest optical strength available, immersed in myosin water solutions with a polymer concentration of $5\,$mg/ml, and $10\,$mM MgCl$_2$, and a) porcine cardiac (PC) myosin (8 curves), b) PC myosin with $10\,$mM ATP (6 curves), c) rabbit skeletal (RS) myosin (12 curves), and d) RS myosin with $10\,$mM ATP (10 curves). The MSDs appear dispersed because we plot jointly all the curves obtained from different trapped probes.}
\end{figure}

To retrieve information of the biopolymer fluctuations on the biological scale \cite{Atakhorrami2014}, we study one-particle microrheology at high frequencies by using optical trapping interferometry (OTI), a technique based in the motorization of the Brownian motion of optically trapped single micro-particles inside the fluid with nanometric accuracy at the microsecond time-scale \cite{franoschresonances2011}.
The short-time scale assures the access to the high-frequency regime of the AM network micro-mechanics, revealing the single filament dynamics \cite{DominguezGarciaFIBRIN2020, DominguezACTIN12023}.
We use skeletal and cardiac myosin (from rabbit skeletal muscle (RS), and porcine cardiac muscle (PC), respectively) purchased from Cytoskeleton, Inc. (with molecular weigth $M_w \approx 200\,$kDa for the myosin heavy chain), and prepared following established methodologies \cite{Pardee1982, Dweck2010}. Myosin II was dissolved in a high-salt solution [$10\,$mM HEPES pH$\,7.5$, $400\,$mM KCl, $1\,$mM DTT] at room temperature. 
We perform the experiments using cardiac and skeletal myosin with a final concentration of $5\,$mg/ml, and $10\,$mM MgCl$_2$ with and without an equimolar amount of ATP. 
A single bead is trapped in the center of a sample chamber using optical tweezers \cite{ashkinapplications1980} and its movement is recorded by means of an interferometric position detector (OTI) \cite{jeneymonitoring2010} (see Sup. Mat. for further details). 
The probes are melamine resin microbeads (Microparticles, GmbH) with a radius $a=1.47\,\mu$m and density $\rho_p = 1570\,$kg/m$^3$ at a work temperature of $T= 21^\circ$C, providing a relatively high refractive index ($n = 1.68$) and good trapping efficiency in OTI experiments. 
The resin beads are chemically non-active and, therefore, the protein specific bonding when surrounded by biomaterials is minimized \cite{valentinecolloid2004, Samiul2012}. 

In a microrheology tracer experiment, the complex modulus of the surrounding material is $G^*(\omega) = G'(\omega)+iG''(\omega)$, where $G'$ is the storage or elastic modulus and $G''$ is the loss modulus, and it is usually calculated from the measured mean-square displacements MSD$(t)\equiv \left<[r(t)-r(0)]^2\right>$ of the microbeads, where ${r}(t)=\{x(t),y(t)\}$ is the one-dimensional bead trajectory. 
In our experiments, we separately analyze the one-dimensional MSDs from the $x(t)$ and $y(t)$ positions of the trapped beads. 
Fig. \ref{fig1} shows all the measured MSDs for cardiac and skeletal myosin solutions with and without adding ATP. 
The measurements contained in Fig. \ref{fig1} are performed using the 
weakest optical strength available (see Supp. Mat.).
The plateau observed at long times in the MSDs curves is related to the constant component in the elastic modulus, because of the optical restoring force, or by the fluid itself \cite{Tassieri2015, Dominguez2016}.
The MSDs increase their curvature and decrease their value at long times in Figs. \ref{fig1} b) and d), reflecting that the influence of ATP modifies the viscoelastic behavior and the network structure of the fluid.  
To extract the complex modulus, $G^*(\omega)$, from the measured MSDs, we apply the standard Mason-Weitz (MW) approach based on the Generalized Stokes-Einstein relation (GSER) \cite{masonparticle1997} and the Mason's approximation \cite{masonestimating2000}.

\begin{figure}
\begin{center}
\includegraphics[scale=1]{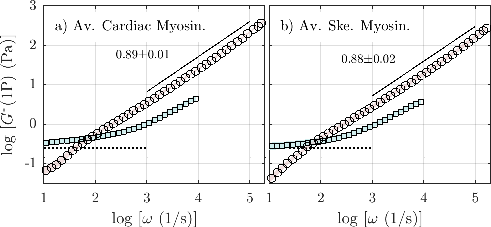}
\end{center}
\caption{\label{fig3}Log-log plots of the averaged loss (\tikzcircle{2pt}) and elastic (\scalebox{0.7}{$\square$}) modulus of myosin solutions with $10\,$mM MgCl$_2$ in the absence of additional ATP when using a) porcine cardiac (PC) myosin, or b) rabbit skeletal (RS) myosin. All data have been blocked in 10 points per decade. Shades represent the standard deviations of the mean. Black lines show the linear regressions to $G''(\omega)$. Dashed lines represent $G'_k = k/6\pi a$, where $k=7\,\mu$m is the stiffness of the optical trap. 
}
\end{figure}

For obtaining a single complex modulus for each case, we have blocked all the MSDs plotted in Fig. \ref{fig1} in 10 points per decade, calculated their complex modulus, and averaged the obtained elastic and loss moduli. 
The result of this procedure is applied first to myosin solutions with $10\,$mM of MgCl$_2$ in absence of ATP. 
In Fig. \ref{fig3}, we plot the averaged $G'(\omega)$ and $G''(\omega)$ as discrete points and their standard deviation of the mean as shades. 
Figs. \ref{fig3} a) and b) show no appreciable differences for the micro-mechanical response of PC and RS myosin without ATP. 
The elastic response of these myosin networks is not particularity strong at low frequencies, because their $G'(\omega)$ value corresponds to the value of the optical trap, $G'_k = k/6\pi a = 0.25\,$Pa. 
The loss modulus, $G''(\omega)$, grows following a power-law behavior $G''(\omega)\sim \omega^\alpha$ at high frequencies. The elastic modulus should behave similarly, but it shows a characteristic breakup at very high frequencies \cite{masonestimating2000, DominguezGarciaFIBRIN2020}. 
Therefore, only $G''(\omega)$ is used to calculate the power-law exponent at high frequencies for both cases. 
Fig. \ref{fig3} shows power-law exponents $\alpha = 0.89(1)$ and $\alpha = 0.88(2)$, compatible with a $7/8$ exponent, which is related to the detection of the longitudinal response of the biopolymers \cite{Everaers1999} 
and has been already observed in fibrin and F-actin experiments using this same experimental technique \cite{DominguezGarciaFIBRIN2020, DominguezACTIN12023}.

\begin{figure}
\begin{center}
\includegraphics[scale=1]{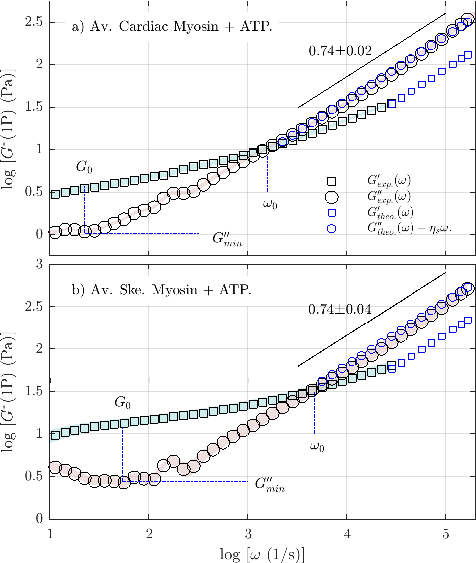}
\end{center}
\caption{\label{fig4}Log-log plots of the averaged loss (\tikzcircle{2pt}) and elastic (\scalebox{0.7}{$\square$}) modulus of myosin solutions with $10\,$mM MgCl$_2$, $10\,$mM ATP, and a) porcine cardiac (PC) myosin, or b) rabbit skeletal (RS) myosin. All data have been blocked in 10 points per decade. Shades represent the standard deviations of the mean. Black lines show the linear regression to the experimental $G''(\omega)$ data. Blue squares and circles represent the theoretical complex modulus by eq. (\ref{ggittes}). 
}
\end{figure}

\begin{table*}
\begin{ruledtabular}
\begin{tabular}{cccc|ccccc}
Myosin+ATP &$\omega_0$ (s$^{-1}$)$\times 10^{3}$ &$G_0$ (Pa) &$G''_{\textrm{min}}$ (Pa) &$l_p$ (nm) &$\zeta$ (Ns/m$^2$)$\times 10^{-3}$ &$\xi$ (nm) &$l_e$ (nm) &$L$ ($\mu$m)\\
\colrule
Cardiac (PC) &$1.6\pm 0.2$ &$2.8 \pm 0.3$ &$1.07 \pm 0.09$ &$68\pm 3$ &$5.89\pm 0.11$ &$113 \pm 4$ &$158 \pm 11$ &$0.42\pm 0.07$\\
Skeletal (RS) &$4.4\pm 0.6$ &$13.6 \pm 1.7$ &$2.7 \pm 0.4$ &$49 \pm 2$ &$7.8\pm 0.2$ &$67 \pm 3$ &$83 \pm 6$ &$0.41 \pm 0.08$
\end{tabular}
\end{ruledtabular}
\caption{\label{tab1} Results obtained by the application of 
eq. (\ref{ggittes}) and the standard theory of polymers 
to the data of Fig. \ref{fig4}.}
\end{table*}

When we repeat this analysis for the PC and RS myosin solutions with added $10\,$mM ATP, we obtain the curves plotted in Fig. \ref{fig4}.
The micro-mechanical behavior detected by the probe has radically changed. The obtained exponents are $\alpha=0.74  \pm 0.02$ for PC and $\alpha=0.74 \pm 0.04$ for RS, and both show a minimum in the loss modulus at lower frequencies,
and a substantial increase of the elastic modulus at lower frequencies.  
Remarkably, the curves for the loss and storage moduli are very similar to those observed for the high-frequency behavior of worm-like micelle solutions \cite{buchananhigh-frequency2005,DominguezGarcia2014}, including the characteristic 3/4 exponent for semiflexible polymers. 

Theoretically, the complex modulus at higher frequencies for a solution of semiflexible polymers is \cite{gittesdynamic1998}:
\begin{equation}
G^{\ast}_{\textrm{theo.}}(\omega) -i\omega\eta_s = \frac{1}{15}\rho_m \,\kappa \, l_p \left(\frac{-2i\zeta}{\kappa}\right)^{3/4} \,\omega^{3/4}\label{ggittes}
\end{equation}
where $\eta_s$ is the solvent viscosity, $\rho_m$ is the polymer concentration in length per unit volume, $\kappa$ is the bending modulus, $l_p$ is the persistence length, and $\zeta$ is the lateral drag coefficient. 
We can evaluate most of these quantities using the standard theory of polymers \cite{doi_theory_1986}. 
The persistence length of the biopolymers is  
$l_p = (k_B T/8\eta_s\omega_0)^{1/3}$, where $k_B$ is the Boltzmann constant, 
$\omega_0$ is the crossover frequency marked in Fig. \ref{fig4},  
which allows to calculate the bending modulus by $\kappa=k_B T \,l_p$.
The lateral drag coefficient, $\zeta$, of the filaments is calculated
\bibnotetext{The lateral drag coefficient is $\zeta=4\pi\eta_s/\ln(A \xi/a)$, where $A$ is a constant which depends of the geometry of the polymer (usually $A=0.6$), and $a = 8\,$nm.} using the mesh size $\xi = (k_B T/G_0)^{1/3}$, where $G_0$ is the value of the elastic modulus at which the loss modulus has a local minimum, denoted as $G''_{\textrm{min}}$ (marked in Fig. \ref{fig4}). 
We need to estimate $\rho_m$, which we calculate as a free parameter to fit the experimental curves.
We obtain $2.0 \times 10^{15}$ m$^{-2}$ for cardiac myosin and $4.2\times 10^{15}$ m$^{-2}$ for skeletal myosin.
Finally, to characterize the structure of the networks, we calculate the entanglement length of the ATP-added myosin networks, $l_e=\xi^{5/3}/l_p^{2/3}$, and the contour length of their filaments, $L = l_e \,G_0/G''_{\textrm{min}}$. 
The results of the calculations for all these quantities are summarized in Table \ref{tab1} and the evaluation of the theoretical complex modulus is plotted in Fig. \ref{fig4} with blue symbols. 
The agreement with the experimental loss modulus in both cases is notable and indicates how the elastic modulus should grow to follow the 3/4 power law behavior and remain significantly smaller than $G''(\omega)$. If the complex viscosity is $\eta \equiv \left|\eta^*(\omega)\right| = \omega^{-1}\,\left[G'(\omega)^2 + G''(\omega)^2\right]^{1/2}$, 
this observation implies $\eta \sim G''(\omega)/\omega$ at high frequencies.

These values allow to evaluate the micro-mechanical behavior of the myosin structures under ATP. 
For example, the persistence length is considered a measure of the degree of flexibility of the structures and the bending force constant evaluates the resistance to bending. The values obtained for $l_p$ for these myosin filaments are similar to those measured for DNA ($\sim 50\,$nm) \cite{Hagerman1981} or worm-like micelles ($\sim 30\,$nm) \cite{Willenbacher_broad_2007,DominguezGarcia2014}, but they are half the value for myosin rods without added ATP ($l_p \sim 140\,$nm) \cite{Hvidt1984}. 
In relation with the bending force, the fundamental flexural mode of a semiflexible rod is characterized by a time\cite{Ookubo1976} $\tau_F = 5.53 \times 10^{-3}\, (\pi \eta_s L^4)/\kappa \ln(L/d)$, where $\eta_s$ is the solvent viscosity, $L$ is the contour length and $d$ the diameter of the rod. 
Using these calculated values and their uncertainties, we obtain an associated frequency interval of $\omega_F \sim 10^3\,$s. This value can be identified with the values of Table \ref{tab1} for the crossover frequency, $\omega_0$, which is the inverse of a relaxation time scale of elastic structures of the material.
The obtained value for the contour length of these filaments is similar for cardiac and skeletal, $L \sim 0.4\,\mu$m, 
matching the expected averaged length of myosin minifilaments \cite{Reisler1980, Trybus1987}.

Therefore, the effect of ATP is to render polymeric structures less rigid and more flexible.
The results of Table \ref{tab1} allow a comparison of the morphology of these filaments with worm-like micelles, already studied using this same experimental procedure \cite{DominguezGarcia2014}. The characteristic lengths are 2-3 times bigger for myosin, but the proportion $L/l_p \sim 6 $ in the case of cardiac myosin is very similar to micelles ($L = 180\,$nm, $l_p = 30\,$nm) and a little higher for skeletal myosin $L/l_p \sim 8$. This factor indicates a slightly coil-like structure in the filaments \cite{Morse_M1_1998}, similar to worm-like micelles, but composed by longer filaments. These coil-like structures have a tendency to bending themselves because of the thermal fluctuations at their independent ends.  
The similarity of the viscoelastic properties of these myosin solutions with worm-like micelles solutions may have its origin in the behavior of the detached myosin heads in the filaments. 
The ATP can dissociate the myosin filaments \cite{Reisler1980}, creating a bipolar structure with heads at both extremes of the filaments. Because of the energy provided by ATP, the myosin heads look for an actin binding site by Brownian search, generating a collective movement with high degree of freedom \cite{Karagiannis2014}. 

Further in-depth, several values in Table \ref{tab1} contain appreciable discrepancies for skeletal and cardiac myosin. 
For example, the stress relaxation time (inverse of $\omega_0$) for cardiac myosin is three times larger than for the skeletal one.  
The persistence length, which is proportional to the bending modulus, is also significatively larger in the activated cardiac myosin, but the elastic modulus at low frequencies, $G'_0$, is five times lower. 
A considerable difference appears in the fitted values for the area density, $\rho_m$, 
where a simple calculation returns the value obtained for cardiac myosin (see Supp. Mat.), but it is doubled for skeletal. 
This outcome is rather complex to interpret, because, in the context for the super-relaxed state, different muscle types show substantial dissimilarities in behavior and distribution of the myosin heads \cite{Hooijman2011}. 
 
These differences between cardiac and skeletal myosin solutions are compatible with a generic mechanical picture of the cardiac and skeletal muscle. 
Skeletal muscle generates more intense forces in comparison with cardiac muscle, which should regularly contract and relax, retaining mechanical capacities \cite{McNamara2015}. In other words, skeletal muscle may remain in a relaxed state more often and has to be able to rapidly generate contraction forces,
but cardiac muscle needs to contract and relax more constantly, normally at a steady pacing.
Cardiac muscle's longer relaxation time supports sustained contractions, conserves energy, ensures rhythmic beating, and allows adaptation to changing demands, protecting the heart and aligning with the super-relaxed state properties of cardiac myosin.

In conclusion, both cardiac and skeletal myosin solutions increase their high-frequency micro-viscoelastic response under the effect of ATP, transforming the chemical energy into mechanical energy. 
This process may be related to the relaxed states, 
in which a mechanism for storing energy prior to muscle contraction is needed. 
Within a similar generic micro-mechanical response, cardiac and skeletal myosin suspensions provide different values regarding their characteristic viscoelastic parameters, something which may be related to several factors such as diffferent MgATPase hydrolysis rates of cardiac and skeletal myosins, distinct cooperative behavior of the myosin heads, or dissimilar physiological functions of cardiac and skeletal muscle. 
\textit{In vitro} studies like this one may be of interest to develop materials whose micro-viscoelasticity can be modified by internal processes, such as energy release or enzymatic activity, with potential 
application in engineering biological materials \cite{Meyers2011}, and in medicine and biotechnology \cite{Joyner2020}. 

See supplementary material for additional information about experimental data and methodology.
P.D.G acknowledges support aid by grant PID2020-117080RB-C54 funded by MCIN/AEI/10.13039/501100011033, and J. C. G\'omez-S\'aez for her proofreading of the texts. J.R.P. acknowledges support from National Institutes of Health grant R01 HL128683. A.A. acknowledges funding from the Swiss National Science Foundation through project PP00P2\_202661. 

%

\end{document}